\documentclass[12pt]{article}
\usepackage{graphicx}
\usepackage{graphics}
\usepackage{amsthm}
\usepackage{amssymb, amsmath}
\title{Analytic results for spatial coherence and information entropy of an optical
vortex field}
\author{Mark W. Coffey\\
Department of Physics\\
Colorado School of Mines\\
Golden, CO  80401\\
USA\\
mcoffey@mines.edu}
\date{July 26, 2014} 
\begin{document}
\maketitle
\baselineskip=25 pt
\begin{abstract}

Optical vortex fields have applications in information processing and storage and in the
manipulation of microscopic particles. We present analytic results for quantities describing
the extent of spatial coherence and entropy of one-dimensional projections of a vortex field. 
Sums of squares of values of certain Jacobi polynomials are important in the analysis.
A family of summation identities is presented that provides the moments of an associated discrete
probability distribution.

\end{abstract}
 
\vspace{.25cm}
\baselineskip=15pt
\centerline{\bf Key words and phrases}
\medskip 
vortex, two-point correlation function, information entropy, hypergeometric function, Jacobi polynomial, Gegenbauer polynomial, associated Legendre polynomial, discrete probability distribution, moments

\bigskip
\bigskip
\noindent
{\bf 2010 PACS codes}
\newline{02.30.Gp, 42.50.Tx} 

\baselineskip=25pt

\pagebreak

\bigskip

Vortices in optical electromagnetic fields have several applications including
information storage and processing and in nanoscience including the manipulation of
microscopic particles.  Recently further experimental demonstrations of the generation of
such vortices have appeared \cite{kumar,mourka,padgettetal,savchenko}. A pair of cylindrical
lenses can be used to convert a laser-generated Hermite-Gaussian (HG) mode into a Laguerre-Gaussian (LG) mode carrying orbital angular momentum \cite{padgettetal}.  Otherwise a common method of creating helical beams is to employ numerically computed holograms.  Fused 
silica whispering gallery mode resonators have been used to generate Bessel beams with 
angular momentum over a thousand $\hbar$ \cite{savchenko}. The numerical work \cite{denicola} used tomographic entropy to characterize the beam profile of LG modes. 

The spatial coherence and information entropy of vortices are of particular importance, and we provide analytic results on these subjects. While certain results have been known for some time \cite{agarwal,beij}, we are able to be more explicit.  In particular, concerning the connnection
coefficients between LG and HG modes, we provide a closed form in terms of the Gauss hypergeometric function $_2F_1$, or equivalently, in terms of certain values of Jacobi polynomials $P_n^{(\alpha,\beta)}(x)$.  These coefficients in turn are used to express a two-point correlation function $\Gamma_{nm}(x,x')$, information entropy $I_{nm}$, and a field purity measure $\mu_{nm}$.
Our analytic study of the entropy in particular has led us to conjecture and prove an identity for the sum of certain values of squares of Jacobi polynomials. 
In fact, we demonstrate a family of summation identities which provides the moments of the corresponding
discrete probability distribution.

We let $L_n^\alpha(x)$ denote the $n$th Laguerre polynomial, $H_n(x)$ the $n$th Hermite polynomial,
$(z)_n=\Gamma(z+n)/\Gamma(z)$ the Pochhammer symbol, and $\Gamma$ the Gamma function
(e.g., \cite{nbs,andrews,grad,rainville}).  
We recall that Hermite polynomials are special cases of Laguerre polynomials, such that
$H_{2n}(x)=(-1)^n 2^{2n}n!L_n^{-1/2}(x^2)$ and $H_{2n+1}(x)=(-1)^n 2^{2n+1}n!xL_n^{1/2}(x^2)$
(e.g., \cite{grad}, p. 1037).
An optical vortex of order $\ell$, carrying angular momentum $\ell \hbar$, has a field
distribution of the separated form $F(r)e^{i\ell \phi}$, such that $F(r)\to 0$ as $r \to 0$.
The waist-plane amplitude of HG modes may be written as
$$u^{HG}_{n,m}(x,y)=\Phi_n(x)\Phi_m(y), \eqno(1)$$
where
$$\Phi_n(x)=\left({\sqrt{2} \over {\sqrt{\pi}2^n wn!}}\right)^{1/2}H_n\left(\sqrt{2}{x \over w}\right)e^{-x^2/w^2}, \eqno(2)$$
with beam waist $w$, and it follows that
$$\int_{-\infty}^\infty \Phi_n(x)\Phi_m(x)dx=\delta_{nm}. \eqno(3)$$
The values $\Phi_n(0)$ may be found by knowing that $H_n(0)$ is $0$ if $n$ is odd, and
$H_n(0)=(-1)^{n/2}n!/(n/2)!$ when $n$ is even.
The relation between LG and HG modes may be written as
$$u^{LG}_{n,m}(x,y)=\sum_{k=0}^{m+n}i^k b(n,m,k)u^{HG}_{m+n-k,k}(x,y). \eqno(4)$$
The connection coefficients $b$ are known, and may be expressed as \cite{beij}
$$b(n,m,k)=\left[{{(n+m-k)!k!} \over {2^{n+m}n!m!}}\right]^{1/2}{1 \over {k!}}{d^k \over {dt^k}}
[(1-t)^n(1+t)^m]_{t=0}. \eqno(5)$$
As we briefly prove below, it follows that 
$$\sum_{k=0}^{m+n}|b(n,m,k)|^2=\sum_{j=0}^{m+n}b^2(n,m,m+n-j)=1. \eqno(6)$$

Defining the two-point correlation function of the one-dimensional projection of an LG beam as
$$\Gamma_{nm}(x,x')=\int_{-\infty}^\infty u^{*LG}_{n,m}(x,y)u^{LG}_{n,m}(x',y)dy, \eqno(7)$$
it follows from (3) and (4) that
$$\Gamma_{nm}(x,x')=\sum_{j=0}^{m+n}b^2(n,m,m+n-j)\Phi_j(x)\Phi_j(x'), \eqno(8)$$
i.e., Mercer's expansion \cite{fock} holds.

{\bf Corollary 1}.  The normalizing sum (6) holds.

{\it Proof}.  Since $\int_{\mathbb{R}^2}|u^{LG}_{n,m}(x,y)|^2dxdy=1$, it follows from (8) that
$$\int_{-\infty}^\infty \Gamma_{nm}(x,x)dx=\sum_{j=0}^{m+n}b^2(n,m,m+n-j)\int_{-\infty}^\infty 
\Phi_j^2(x)dx=\sum_{j=0}^{m+n}b^2(n,m,m+n-j)=1.$$
\qed

{\bf Lemma 1}.
$${d^k \over {dt^k}}[(1-t)^n(1+t)^m]_{t=0}={{m!} \over {(m-k)!}} ~_2F_1(-k,-n;m-k+1;-1).  
\eqno(9)$$

{\it Proof}.  First note that
$${d^k \over {dt^k}}(1+at)^n={{n!} \over {(n-k)!}}a^k (1+at)^{n-k}, ~~~~0\leq k\leq n, \eqno(10)$$
and is $0$ for $k>n$.  It follows from the product rule that
$${d^k \over {dt^k}}[(1-t)^n(1+t)^m]=\sum_{\ell=0}^k{k \choose \ell}\left({d \over {dt}}\right)^{k-\ell}(1+t)^m \left({d \over {dt}}\right)^\ell(1-t)^n$$
$$=\sum_{\ell=0}^k (-1)^\ell {k \choose \ell}{{m!} \over {(m-k+\ell)!}}{{n!} \over {(n-\ell)!}}
(1+t)^{m-k+\ell}(1-t)^{n-\ell}.$$
Therefore, repeatedly using the relation $n!/(n-k)!=(-1)^k (-n)_k$, and the series definition of
$_2F_1$, we deduce Lemma 1.  \qed 

{\bf Corollary 2}.  We have from (5) and (9)
$$b^2(n,m,k)={{(n+m-k)!m!} \over {n!2^{n+m}k!}} {1 \over {[(m-k)!]^2}} ~_2F_1^2(-k,-n;m-k+1;-1),
\eqno(11)$$
and from (6) and (10) the summation identity
$$\sum_{j=0}^{m+n}b^2(n,m,m+n-j)={{m!} \over {2^{n+m}n!}}\sum_{j=0}^{m+n} {{j!} \over {(m+n-j)!}}
{1 \over {[(j-n)!]^2}} ~_2F_1^2(j-m-n,-n;j-n+1;-1)=1. \eqno(12)$$

An alternative form of the coefficients $b^2(n,m,k)$ may be obtained via Jacobi polynomials
$P_n^{(\alpha,\beta)}(x)$, as we have (e.g., \cite{grad}, p. 1035)
$$P_n^{(\alpha,\beta)}(x)={{(-1)^n} \over {2^nn!}}(1-x)^{-\alpha}(1+x)^{-\beta}{d^n \over {dx^n}}
[(1-x)^{\alpha+n}(1+x)^{\beta+n}].  \eqno(13)$$
We find that
$$b(n,m,k)=\sqrt{{(n+m-k)!k!} \over {2^{n+m}n!m!}}2^k (-1)^k P_k^{(n-k,m-k)}(0)$$
$$=\sqrt{{(n+m-k)!n!} \over {2^{n+m}k!m!}}{{(-1)^k} \over {(n-k)!}} ~_2F_1(-k,-m;n-k+1;-1), \eqno(14)$$
wherein we have used (\cite{grad}, p. 1036).  Other forms of $b(n,m,k)$ in terms of $_2F_1(1/2)$ or
$_2F_1(-1)$ are easily written (\cite{grad}, p. 1036). According to Corollary 1 we then have the identity
$$\sum_{k=0}^{m+n} b^2(n,m,k)={1 \over {2^{n+m}n!m!}}\sum_{k=0}^{m+n} (n+m-k)!k! 2^{2k} [P_k^{(n-k,
m-k)}(0)]^2=1.  \eqno(15)$$
With the summation over both lower and upper indices of $P_k^{(n-k,m-k)}(0)$ in (15), this
sum appears to be different from those studied by Dette with squares of Jacobi polynomials
(\cite{dette}, section 5).

Given the normalizing sum of Corollary 1, the coefficients $b^2$ may be taken as probabilities
and an entropy of the Shannon-Boltzmann-Gibbs type defined,
$$I_{nm}=-\sum_{j=0}^{m+n}b^2(n,m,m+n-j)\ln b^2(n,m,m+n-j).  \eqno(16)$$
Then from the first expression in (14) we have
$$I_{nm}=(m+n)\ln 2+\ln n!+\ln m!$$
$$-\sum_{k=0}^{m+n} {{(n+m-k)!k!} \over {2^{n+m}n!m!}}2^{2k} [P_k^{(n-k,m-k)}(0)]^2
\ln \left[(n+m-k)!k!2^{2k} [P_k^{(n-k,m-k)}(0)]^2\right].  \eqno(17)$$
As an example, a plot of $I_{n,N-n}$ versus $n$ for $N=20$ is given in Figure 1.
The entropy $I_{n,N-n}$ is not strictly concave downward with $n$.  Indeed, as expected from
\cite{agarwal}, there is a local minimum in this entropy for $n=N/2$ when $N$ is even,
corresponding to a nonvortex state.  

The discrete probability distribution $b^2(25,25,k)$ is plotted versus $k$ in Figure 2
and $b^2(7,25,k)$ similarly in Figure 3.  Two features are evident in the $m=n=25$ case:
there are alternating $0$ values and the distribution is quite flat away from the endpoints.
Therefore, the entropy is expected to be lower in such a situation.  In contrast in Figure 3,
there is but one value of exactly $0$ and there is much more scatter to the distribution, indicating
higher entropy.  Below we calculate the moments of $b^2(n,m,k)$ and from these we find that the
variance $\sigma_{nm}^2=\langle k^2 \rangle-\langle k \rangle^2= (2mn+m+n)/4$, where $\langle 
\rangle$ denotes expectation. The variance $\sigma^2_{n,25-n}$ is plotted in Figure 4 and
it is maximal at $[(n+m)/2]$.  The larger variance at $n=m=[N/2]$ is also an indicator that
smaller entropy could result.  The skewness $\langle(k-\langle k \rangle)^3\rangle =\langle k^3
\rangle -3\langle k^2\rangle \langle k\rangle+2\langle k\rangle^3$ may be computed from 
(18), (20), and (21) below.  As expected, it evaluates to $0$.   

Further analytic progress from (17) is possible, as we now show.  We first made, and then proved,
the following.
{\newline \bf Conjecture 1}.  
$$-2\ln 2\sum_{k=1}^{m+n} {{(n+m-k)!k!} \over {2^{n+m}n!m!}}2^{2k} k[P_k^{(n-k,m-k)}(0)]^2
=-(\ln 2)(m+n).  \eqno(18)$$
Note that since this conjecture holds, the contribution $(m+n)\ln 2$ in the first line of (17) is
precisely cancelled.  
Furthermore, we have the equality of contributions from (17)
$$-\sum_{k=0}^{m+n} {{(n+m-k)!k!} \over {2^{n+m}n!m!}}2^{2k} [P_k^{(n-k,m-k)}(0)]^2
\ln (n+m-k)!$$
$$= -\sum_{k=0}^{m+n} {{(n+m-k)!k!} \over {2^{n+m}n!m!}}2^{2k} [P_k^{(n-k,m-k)}(0)]^2
\ln k!.  \eqno(19)$$

The identity (18) is part of a family of summation relations, as we also indicate.  
For example, we suspected that the sum
$$2\sum_{k=1}^{m+n} {{(n+m-k)!k!} \over {2^{n+m}n!m!}}2^{2k} k^2[P_k^{(n-k,m-k)}(0)]^2 
={1 \over 2}(m^2+n^2)+2mn+{1 \over 2}(m+n) \eqno(20)$$
is expressible as a quadratic function of $n$ and $m$, as is indeed the case.  Furthermore,
$$4\sum_{k=1}^{m+n} {{(n+m-k)!k!} \over {2^{n+m}n!m!}}2^{2k} k^3[P_k^{(n-k,m-k)}(0)]^2 
={1 \over 2}(m^3+n^3+9m^2n+9mn^2)+{3 \over 2}(m^2+n^2)+3mn, \eqno(21)$$
$$2\sum_{k=1}^{m+n} {{(n+m-k)!k!} \over {2^{n+m}n!m!}}2^{2k} k^4[P_k^{(n-k,m-k)}(0)]^2$$
$$={1 \over 8}(m^4+n^4)+2(m^3n+mn^3)+{9 \over 2}m^2n^2+{3 \over 4}(m^3+n^3)+3(m^2n+mn^2)
+{3 \over 8}(m^2+n^2)-{1\over 2}mn-{1 \over 4}(m+n), \eqno(22)$$
$$4\sum_{k=1}^{m+n} {{(n+m-k)!k!} \over {2^{n+m}n!m!}}2^{2k} k^5[P_k^{(n-k,m-k)}(0)]^2$$
$$={1 \over 8}(m^5+n^5)+{{25} \over 8}(m^4n+mn^4)+{{25} \over 2}(m^3n^2+m^2n^3)+{5 \over 4}(m^4+n^4)+{{35} \over 4}(m^3n+mn^3)$$
$$+15m^2n^2+{{15} \over 8}(m^3+n^3)-{5\over 8}(m^2n+mn^2)
-{5 \over 4}(m^2+n^2)+{5 \over 2}mn, \eqno(23)$$
and
$$4\sum_{k=1}^{m+n} {{(n+m-k)!k!} \over {2^{n+m}n!m!}}2^{2k} k^6[P_k^{(n-k,m-k)}(0)]^2$$
$$={1 \over {16}}(m^6+n^6)+{9 \over 4}(m^5n+mn^5)+{{225} \over {16}}(m^4n^2+m^2n^4)+25m^3n^3+{{225} \over 8}(m^3n^2+m^2n^3)$$
$$+{{15} \over {16}}(m^5+n^5)+{{165} \over {16}}(m^4n+mn^4)+{{45} \over {16}}(m^4+n^4)
+{{35} \over 8}(m^3n+mn^3)-{{45} \over 8}m^2n^2$$
$$-{{15} \over {16}}(m^3+n^3)-{{135} \over {16}}(m^2n+mn^2)-{{15} \over 8}(m^2+n^2)+{{13} \over 4}mn+m+n
\eqno(24)$$

Let us note the special case connection of Gegenbauer polynomials $C_n^\lambda(x)$ with Jacobi polynomials (\cite{grad}, p. 1036, \cite{andrews}, p. 302),
$$C_n^\lambda(x)={{(2\lambda)_n} \over {(\lambda+1/2)_n}} P_n^{(\lambda-1/2,\lambda-1/2)}(x).
\eqno(25)$$
These polynomials are orthogonal on $[-1,1]$ with weight function $(1-x^2)^{\lambda-1/2}$. 
Then various sums of this paper including (15)-(23) may be rewritten when $n=m$ with the relation
$$P_k^{(n-k,n-k)}(0)={{(n-k+1)_k} \over {(2n-2k+1)_k}}C_k^{n-k+1/2}(0)
={{n!} \over {(n-k)!}}{{(2n-2k)!} \over {(2n-k)!}}C_k^{n-k+1/2}(0).  \eqno(26)$$
As an illustration, when $n=m$, (20) takes the form
$$2\sum_{k=1}^{2n} {{4^{k-n}k^2k!} \over {(2n-k)!}}\left[{{(2n-2k)!} \over {(n-k)!}}\right]^2
[C_k^{n-k+1/2}(0)]^2 = (3n+1)n. \eqno(27)$$
With the aid of the duplication formula for the Gamma function,
$${{(2n-2k)!} \over {(n-k)!}}={4^{n-k} \over \sqrt{\pi}}\Gamma\left(n-k+{1 \over 2}\right),
\eqno(28)$$
in this case (20) may also be rewritten as
$${2 \over \pi}\sum_{k=1}^{2n}{{4^{n-k}k^2k!} \over {(2n-k)!}}\Gamma^2\left(n-k+{1 \over 2}\right)
[C_k^{n-k+1/2}(0)]^2 = (3n+1)n. \eqno(29)$$
When considering this sum inductively, the following is a useful property:
$C_n^{\lambda+1}(0)=[1+n/(2\lambda)]C_n^\lambda(0)$.
We have the values $C_{2n+1}^\lambda(0)=0$ and 
$$C_{2n}^\lambda(0)={{(-1)^n} \over {(\lambda+n)B(\lambda,n+1)}}={{(-1)^n} \over {n!}}
(\lambda)_n={{-\lambda} \choose n},  \eqno(30)$$
wherein $B(x,y)=\Gamma(x)\Gamma(y)/\Gamma(x+y)$ is the Beta function.  Then the summation of
(29) becomes
$${8 \over \pi}\sum_{m=1}^n{{4^{n-2m}m^2 (2m)!} \over {(2n-2m)!}}\Gamma^2\left(n-2m+{1 \over 2}\right)[C_{2m}^{n-2m+1/2}(0)]^2$$ 
$$={8 \over \pi}\sum_{m=1}^n{{4^{n-2m}m^2 (2m)!} \over {(2n-2m)!}}\Gamma^2\left(n-2m+{1 \over 2}\right){{2m-n-1/2} \choose m}^2$$
$$=8(-1)^n \sum_{m=1}^n {{(-1)^m m \Gamma(m+1/2)} \over {(m-1)!(n-m)!\Gamma(m-n+1/2)}} 
= (3n+1)n. \eqno(31)$$
Previous to the last equality, we have twice used both $\Gamma(z)\Gamma(1-z)=\pi/\sin \pi
z$ and the duplication formula of the Gamma function, with a result that may easily be
written in terms of Pochhammer symbols.  
The summation (31) is a special case of 
$${8 \over \pi}\sum_{m=1}^n{{4^{n-2m}m^2 (2m)!} \over {(2n-2m)!}}\Gamma^2\left(n-2m+{1 \over 2}\right){{2m-n-1/2} \choose m}^2 z^m$$
$$= {{4z\Gamma(n-1/2)} \over {\sqrt{\pi}(n-1)!}} ~_3F_2\left(2,{3 \over 2},1-n;1,{3 \over 2}-n;z
\right), \eqno(32)$$
where $_3F_2$ is a generalized hypergeometric function.
It is a simple matter to show the reduction
$$~_3F_2\left(2,{3 \over 2},1-n;1,{3 \over 2}-n;z\right)=~_2F_1\left({3 \over 2},1-n;{3 \over 2}
-n;z\right)+{3 \over 2}{{(n-1)} \over {(n-3/2)}} z~_2F_1\left({5 \over 2},2-n;{5 \over 2}
-n;z\right).$$
Then when $z=1$ one can show that
$$~_3F_2\left(2,{3 \over 2},1-n;1,{3 \over 2}-n;1\right)={{n!} \over \sqrt{\pi}}(-1)^{n-1}
{1 \over 4}(3n+1)\Gamma\left({3 \over 2}-n\right).$$
Then from (32) the summation result of (31) again follows.

More generally, we have the family of summations
$${8 \over \pi}\sum_{m=1}^n {{m^k (2m)!4^{n-2m}} \over {(2n-2m)!(m!)^2}} \Gamma^2(n-m+1/2)z^m$$
$$={{4z\Gamma(n-1/2)} \over {\sqrt{\pi}(n-1)!}} ~_{k+1}F_k\left({3 \over 2},1-n,2,\ldots,2;{3 \over 2}-n,1,\ldots,1;z\right),$$
with $_pF_q$ the generalized hypergeometric function.  The functions $~_{k+1}F_k(z)$ may 
again be reduced in terms of a sum of Gauss hypergeometric functions.  For example,
$$~_4F_3\left(2,2,{3 \over 2},1-n;1,1,{3 \over 2}-n;z\right)=~_2F_1\left({3 \over 2},1-n;{3 \over 2}-n;z\right)+{9 \over 2}{{(n-1)} \over {(n-3/2)}} z $$
$$ \times ~_2F_1\left({5 \over 2},2-n;{5 \over 2}-n;z\right)
+{{15} \over 4}{{(n-1)} \over {(n-3/2)}}{{(n-2)} \over {(n-5/2)}} z^2~_2F_1\left({7 \over 2},3-n;{7 \over 2}-n;z\right).$$

We mention that such $_2F_1(z)$ functions, useful in obtaining the moments of the discrete
distribution $b^2(n,n,k)$, may be found from derivatives of the Legendre polynomial $P_n$.
In particular,
$${}_2F_1\left({1 \over 2},-n;{1 \over 2}-n;z\right)={{n!z^{n/2}} \over {(1/2)_n}} P_n\left({{1+z} \over 
{2\sqrt{z}}}\right).$$
This formula is a transformation of other hypergeometric forms of $P_n(x)$ \cite{rainville} (p. 164).
Since $(d/dz) {}_2F_1(a,b;c;z)=(ab/c) {}_2F_1(a+1,b+1;c+1;z)$, the moments may be determined by
differentiation followed by putting $z=1$.

The associated Legendre polynomials $P_n^m(x)$ (\cite{andrews}, p. 456, \cite{grad}, p. 1031) are related to Gegenbauer polynomials via
$$C_{n-m}^{m+1/2}(x)={1 \over {(2m-1)!!}} {{d^mP_n(x)} \over {dx^m}}=(-1)^m 2^m {{m!}\over 
{(2m)!}}(1-x^2)^{-m/2} P_n^m(x), \eqno(33)$$
wherein $P_n(x)=P_n^0(x)$ are the ordinary Legendre polynomials.  Therefore, another way to 
write (29) is
$$2 \sum_{k=1}^{2n} {{k^2 k!} \over {(2n-k)!}} [P_n^{n-k}(0)]^2=(3n+1)n.  \eqno(34)$$
Similarly, other sums such as (15)-(19) and (21)-(24) may be expressed in terms of $P_n^{n-k}(0)$
when $n=m$.  The Appendix discusses the normalizing sum (15) when $n=m$ in terms of associated Legendre polynomials.  With the values given in (A.4) it is easily shown that (34) is 
equivalent to (31), and to (32) with $z=1$.

It is important for the following considerations to state this result:
$$\sum_{k=0}^\infty {{k!} \over {(-a-b)_k}} t^k P_k^{(a-k,b-k)}(x)P_k^{(a-k,b-k)}(y)
=\left[1-(x+1)(y+1){t \over 4}\right]^a \left[1-(x-1)(y-1){t \over 4}\right]^b$$
$$\times _2F_1\left\{-a,-b;-a-b;-t\left[1-(x+1)(y+1){t \over 4}\right]^{-1}\left[1-(x-1)(y-1)
{t \over 4}\right]^{-1}\right\}. \eqno(35)$$
Manocha employed a two-variable hypergeometric function $F_2(a;b,b';g,g';x,y)$ and 
demonstrated a generalization of (35) \cite{manocha}.  When $m=0$ in (4) of that reference,
(9) there results, as we have given for (35).  This case may also be found in \cite{manocha2} (15),
being first proved by Carlitz \cite{carlitz}.

We first note that (15) may be obtained as a special case of (35), rewriting (15) as
$$\sum_{k=0}^{m+n} b^2(n,m,k)={1 \over 2^{n+m}}{{n+m} \choose m} \sum_{k=0}^{m+n}{{k! (-1)^k} \over {(-m-n)_k}} 4^k [P_k^{(n-k,m-k)}(0)]^2. \eqno(36)$$
Now we put $x=y=0$ and $t=-4$ in (35) to find
$$\sum_{k=0}^{a+b}{{k!} \over {(-a-b)_k}}(-4)^k [P_k^{(a-k,b-k)}(0)]^2
=2^{a+b} ~_2F_1(-a,-b;-a-b;1)$$
$$=2^{a+b}{{a+b} \choose a}^{-1}.  \eqno(37)$$
When $a$ and $b$ are integers, the sum truncates for $k>a+b$ in (35), and this verifies (once more) (15).

{\it Proof} of (18).  We rewrite the needed sum as
$$\sum_{k=1}^{m+n} {{(n+m-k)!k!} \over {2^{n+m}n!m!}}2^{2k} k[P_k^{(n-k,m-k)}(0)]^2
={{{n+m} \choose m} \over 2^{n+m}}\sum_{k=1}^{m+n}{{k! k (-4)^k} \over {(-m-n)_k}} [P_k^{(n-k,m-k)}(0)]^2. \eqno(38)$$
We put $x=y=0$ in (35) and differentiate with respect to $t$.  We then
put $t=-4$.  So we start with the function
$$f(t)=\left(1-{t \over 4}\right)^{a+b} ~_2F_1\left[-a,-b;-a-b;-{t \over {(1-t/4)^2}}\right]
=\sum_{k=0}^\infty {{(-t)^k} \over {{{a+b} \choose k}}} [P_k^{(a-k,b-k)}(0)]^2.  \eqno(39)$$
Differentiating with respect to $t$ with the product rule, the needed nonzero term when $t\to -4$
is
$$-{1 \over 4}(a+b)\left(1-{t \over 4}\right)^{a+b-1} ~_2F_1\left[-a,-b;-a-b;-{t \over {(1-t/4)^2}}\right].$$
Upon putting $t=-4$, this quantity becomes
$$-{1 \over 4}(a+b)2^{a+b-1}{{a+b} \choose a}^{-1},$$
from which (18) follows.  \qed

The proof of (20) is similar:  we differentiate $t df/dt$ with respect to $t$ and then 
put $t=-4$.  
For that calculation is it useful to to recall the values $_2F_1(1-a,1-b;1-a-b;1)=a!b!/(a+b-1)!$.
Now that results such as (20) and the summation identities below it have been stated, they may also be proved by induction.

Other forms of the entropy may be used, including the Renyi form with parameter $\alpha$,
$$I_{nm}^\alpha = {1 \over {1-\alpha}}\ln \sum_{k=0}^{m+n} b^{2\alpha}(n,m,k).  \eqno(40)$$
A plot of $I_{nN-n}^\alpha$ versus with $N=20$ and $\alpha=2$ is given in Figure 5.  Like
$I_{nN-n}$, this entropy is not strictly concave downward as a function of $n$.  However,
the local minimum at $N/2$ when $N$ is even is much less pronounced than for $I_{nN-n}$.

A measure of the purity of an optical field is
$$\mu_{nm}={{\int \int |\Gamma_{nm}(x,x')|^2dxdx'} \over {[\int \Gamma_{nm}(x,x)dx]^2}}
={{\sum_{j=0}^{m+n}b^4(n,m,m+n-j)} \over {[\sum_{j=0}^{m+n}b^2(n,m,m+n-j)]^2}}$$
$$=\sum_{j=0}^{m+n}b^4(n,m,m+n-j), \eqno(41)$$
wherein we have applied Corollary 1.  A fully coherent beam has $\mu_{nm}=1$ and a partially 
coherent beam has $0<\mu_{nm}<1$. Again, our analytic expressions for $b(n,m,k)$ may be
used in evaluating this field purity measure.

The Wigner function is related to the Fourier transform of the two-point correlation function.
As such, the squared coefficients $b^2(n,m,k)$ will again appear in summations, and our
analytic expressions used.


\bigskip
\centerline{\bf Appendix:  The normalizing sum for $b^2(n,n,k)$}
\centerline{\bf in terms of associated Legendre polynomials $P_n^m(x)$}

When $n=m$, the normalizing sum of (15) may be written as
$$\sum_{k=0}^{2n} {{k!} \over {(2n-k)!}}[P_n^{n-k}(0)]^2=1.  \eqno(A.1)$$
When proving this relation inductively, the following property (\cite{grad}, p. 1005) is
key:  $P_\nu^{\mu+1}(0)=-(\nu+\mu)P_{\nu-1}^\mu(0)$.  Then
$$\sum_{k=0}^{2n+2} {{k!} \over {(2n-k+2)!}}[P_{n+1}^{n-k+1}(0)]^2
=\sum_{k=0}^{2n+2} {{k!} \over {(2n-k+2)!}}(2n-k+1)^2 [P_n^{n-k}(0)]^2$$
$$=\sum_{k=0}^{2n} {{k!} \over {(2n-k)!}}{{(2n-k+1)} \over {(2n-k+2)}} [P_n^{n-k}(0)]^2
+(2n+2)![P_n^{-n-2}(0)]^2.  \eqno(A.2)$$
The inductive step follows as a result of the following identity:
$$\sum_{k=0}^{2n}{{k! [P_n^{n-k}(0)]^2} \over {(2n-k)!(2n-k+2)}}=(2n+2)![P_n^{-n-2}(0)]^2
={1 \over 4^{n+1}}{{2n+2} \choose {n+1}}.  \eqno(A.3)$$
The values $P_n^{-n-2}(0)=2^{-n-1}/(n+1)!$ follow from \cite{grad}, p. 1011.

We also record here the following special values:  $P_n^n(0)=(-1)^n(2n-1)!!$ and 
$P_n^{-n-1}(0)=1/(2n+1)!!$, where $(2n+1)!!=(2n+1)(2n-1)\cdots 3$.  In addition, from
\cite{grad}, p. 1011 we have
$$P_n^{n-k}(0)={2^{n-k} \over \sqrt{\pi}}\cos\left[(2n-k){\pi \over 2}\right]{{\Gamma\left(
{{2n-k+1} \over 2}\right)} \over {\Gamma\left({k \over 2}+1\right)}}.  \eqno(A.4)$$

\begin{figure}[h]
\begin{center}
\includegraphics[height=3.25in,width=4.0in,angle=0]{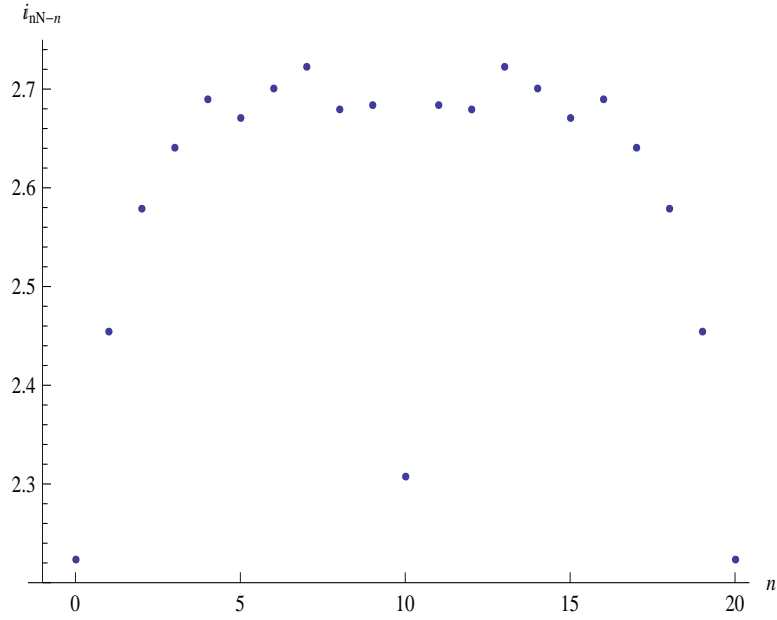}
\caption{Values of the entropy $I_{nN-n}$ for $N=20$ as a function of $n$.}
\end{center}
\end{figure}
  
\begin{figure}[h]
\begin{center}
\includegraphics[height=3.25in,width=4.0in,angle=0]{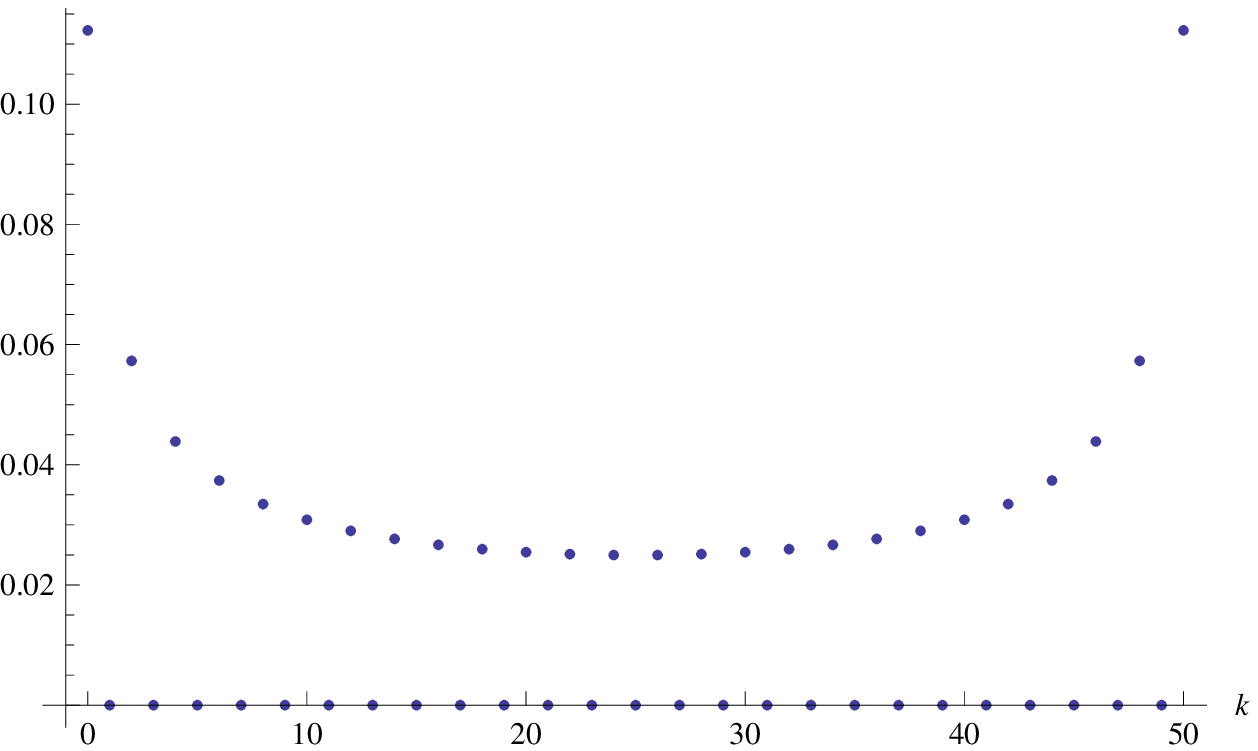}
\caption{Values of the discrete distribution $b^2(25,25,k)$ versus $k$.}
\end{center}
\end{figure}
  
\begin{figure}[h]
\begin{center}
\includegraphics[height=3.25in,width=4.0in,angle=0]{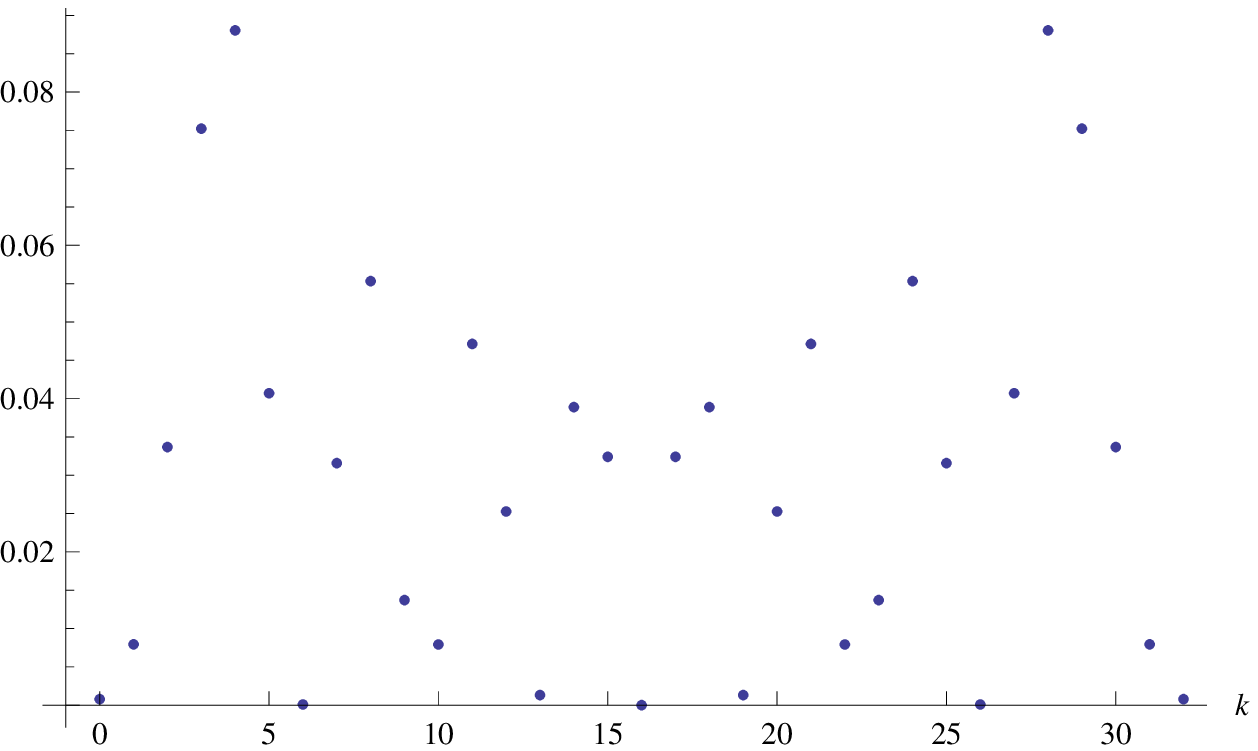}
\caption{Values of the discrete distribution $b^2(7,25,k)$ versus $k$.}
\end{center}
\end{figure}

\begin{figure}[h]
\begin{center}
\includegraphics[height=3.25in,width=4.0in,angle=0]{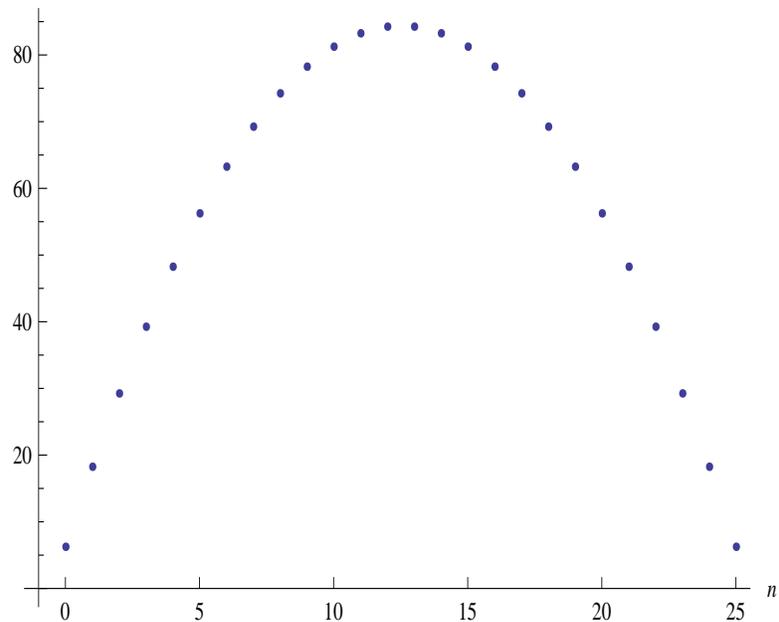}
\caption{Values of the variance $\sigma^2_{nN-n}$ for $N=25$ versus $n$.}
\end{center}
\end{figure}

\begin{figure}[h]
\begin{center}
\includegraphics[height=3.25in,width=4.0in,angle=0]{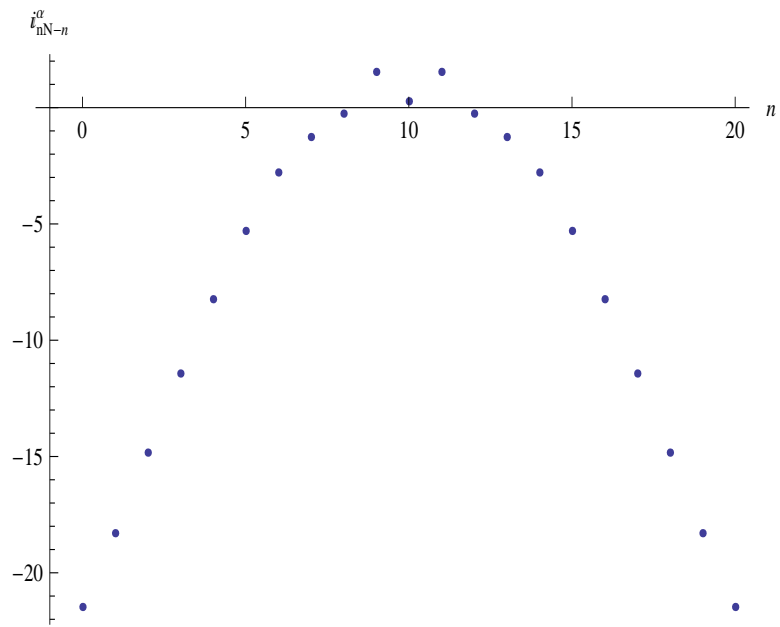}
\caption{Values of the Renyi entropy $I_{nN-n}^\alpha$ for $N=20$ and $\alpha=2$ as a function of $n$.}
\end{center}
\end{figure}

\pagebreak

\end{document}